# Protein and hydration-water dynamics are decoupled: A new model connecting dynamics and biochemical function is required.

Antonio Benedetto[1,2]

[1]School of Physics, University College Dublin, Dublin 4, Ireland
[2]Laboratory for Neutron Scattering, Paul Scherrer Institut, Villigen, Switzerland
antonio.benedetto@ucd.ie

**Abstract:**

Water plays a major role in bio-systems, greatly contributing to determine their structure, stability and even function. It is well know, for instance, that proteins require a minimum amount of water to be functionally active. Since the biological functions of proteins involve changes of conformation, and sometimes chemical reactions, it is natural to expect a connection of these functions with dynamical properties of the coupled system of proteins and their hydration water. However, despite many years of intensive research, the detailed nature of protein – hydration water interactions, and their effect on the biochemical activity of proteins through peculiar dynamical effects, is still partly unknown. In particular, models proposed so far, fail to explain the full set of experimental data. The well-accepted "*protein dynamical transition*" scenario is based on perfect coupling between the dynamics of proteins and of their hydration water, which has never been confuted experimentally. We present high-energy resolution elastic neutron scattering measurements of the atomistic dynamics of the model protein, lysozyme, in water that were carried out on the IN16B spectrometer at the Institut Laue-Langevin in Grenoble, France. These show for the first time that the dynamics of proteins and of their hydration water are actually de-coupled. This important result militates against the well-accepted scenario, and requires a new model to link protein dynamics to the dynamics of its hydration water, and, in turn, to biochemical function.

**Keywords**: neutron scattering, protein hydration water, protein dynamics, protein dynamical transition, protein function, biophysics

In 560 BC Thales hypothesized that water is the primary essence of life[1], and nowadays, this hypothesis has been well accepted. Since a few decades, it has become well established that water molecules actively interact with, and support, the biochemistry of different classes of biomolecules[2-11]. In the case of proteins, for instance, the water molecules adsorbed at the protein surface play a major biological role, since it is well known that proteins require a minimum amount of water to be biologically active. These water molecules form the so-called *protein hydration layers*, usually consisting of one or two layers adsorbed at the protein surface, of thickness ranging between 5 to 10 Å, and accounting for about 20% of the hydrated protein weight (Fig. 1a). Even thermal denaturation, which might also be seen as the analogue of other biological processes such as amyloidogenesis, needs this minimum amount of water to occur (Fig. 1b, cycle 1). These water molecules display physico-chemical properties that are apparently different from molecules in pure bulk water. For instance, they are prevented from crystallising just below 0 C because the competition of their mutual interaction and their hydrogen bonding to the protein make it difficult for them to rearrange into the typical tetrahedral ice structure (Fig. 1b, cycle 1). Remarkably, proteins' denaturation restores their ability to crystallise (Fig. 1b, cycle 2).

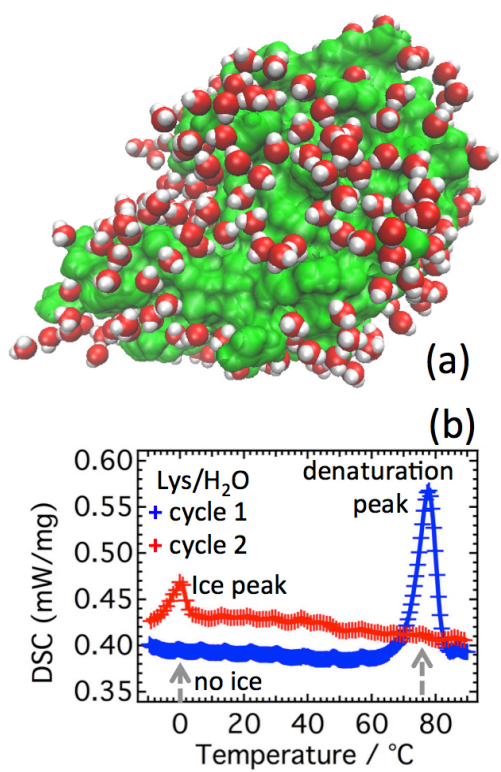

Fig. 1 – Lysozyme and its hydration water. (a) Crystal structure of human lysozyme (green) together with its (about 280) hydration water molecules obtained by X-ray at 1.8 Å of spatial resolution.[9] This is the amount of water molecules required for the protein to be biological active. (b) Differential scan calorimetric heating profiles of hydrated lysozyme before (blue), and after (red) thermal denaturation (blue peak). In hydrated lysozyme, the hydration water due to the interaction with the protein is not able to form ice, but after the thermal denaturation, due to a different state of the protein, part of the hydration water molecules recover their ability to form ice (red peak).

Since the biological functions of proteins involve changes of conformation and sometimes chemical reactions, it is natural to expect a connection of these functions with dynamical properties of the coupled system of proteins with their hydration water. An important thread in this discussion started in 1989 with the seminal neutron scattering work of W. Doster and co-workers[12] pointing to a sharp increase in the atomic mean-square displacement of a hydrated model protein at around $T_{PDT} \approx 220K$. The authors referred to this transition as the "*protein dynamical transition*" (PDT). Because proteins seem to become biologically active around this temperature[13,14], and since this transition was not observed in the absence of hydration water, the authors proposed to link the PDT to the functioning of proteins. The PDT has since been observed in several other proteins, solvated not only by pure water, e.g. in glycerol[15,16] and water solutions of disaccharides[16,17]. The most clear and direct way to measure such PDT experimentally is by *elastic neutron scattering*, which is collected as function of sample-temperature in what is called "fix-windows scan". This plot shows an abrupt decrease for the hydrated protein with respect to the dry one at $T_{PDT}$ (Fig. 2). Consequently, the research community have focussed on understanding the protein-water relaxation mechanisms underlying the PDT and, in turn, the biochemical function.

To the best of our knowledge, the only attempt to propose a unified model of protein dynamics was made by H. Frauenfelder and co-workers[18]. In their paper they proposed that protein motions are modulated by the hydration shell and by the bulk solvent. Essentially, they proposed that: (i) large-scale protein motions are coupled to fluctuations in the bulk solvent that are controlled by the solvent viscosity, which are absent in a solid environment; whereas (ii) internal protein motions are coupled to the beta fluctuations of the hydration shell, they are controlled by hydration, and are absent in dehydrated proteins. This intriguing scenario in which water molecules drive protein dynamics and, in turn, their functions, has to be considered together with the relaxation dynamics of the protein hydration water itself. This was measured by Chen and co-workers[19] in which a crossover from Arrhenius (strong glass-like) to non-Arrhenius (fragile glass-like) behaviour was found at $T_{FSC} \approx 220K$. The picture that emerges is even more intriguing: the change in the hydration-water dynamics at $T_{FSC}$ triggers the protein internal-motions that also became active at $T_{FSC}$. This would mean that at $T_{FSC}$ a sudden variation in the protein dynamics should occur as well. The scenario emerging from this logical implication is also justified by the fact that $T_{PDT} \approx T_{FSC} \approx 220K$.

The link between the transition in the hydration-water dynamics and the PDT has been observed in all neutron-scattering experiments on this subject so far. Again, as above mentioned, the elastic neutron scattering intensity versus temperature represents one of the most used observables to study such protein – hydration water dynamical entanglement, which allows the dynamics of proteins and of their hydration water to be probed independently. To achieve this, two elastic-scattering profiles of the hydrated protein are collected: one by hydrating with $H_2O$, and the other with $D_2O$; in both cases with an amount of water corresponding to the protein hydration water as in Fig. 1a (i.e. about between 0.3 to 0.4g of water per gram of protein). The elastic spectrum of the protein in the dry state has to be also collected to provide a baseline. The basis is that neutrons are very sensitive to hydrogen atoms and much less sensitive to deuterium, enabling a distinction between the isotopes. The biochemical function however, is only slightly affected by the use of heavy water instead of water. In contrast, X-rays are not able to see hydrogen atoms, or distinguish isotopes. As a result, when measuring with neutrons the contribution of $D_2O$ to the signal from a protein hydrated in $D_2O$ is negligible, and relaxations in the elastic spectra can be related to the protein itself (which also contains a lot of hydrogen atoms). In this way it is possible to probe the relaxation dynamics of protein alone, yet hydrated, and extract $T_{PDT}$. In the case of a protein hydrated in $H_2O$, the contributions of water and protein have very similar weights in the signal (i.e. the density of hydrogen atoms in the protein and in hydration water are nearly equivalent). As a result, by measuring the protein hydrated in $H_2O$, relaxations in the elastic spectra arise from relaxations of either the protein or its hydration water. By comparison of the spectra of the protein in $D_2O$ with that in $H_2O$, it is possible to determine the relaxation process of hydration water alone, and, in turn, extract its transition temperature.

With the current state-of-the art, all experiments so far show the proteins and their hydration water to have the same transition temperature, which strongly supports the PDT scenario described above. That is, the dynamics of proteins and of their hydration water are strongly coupled. In Fig. 2 the classic example of the model protein, lysozyme, is shown.

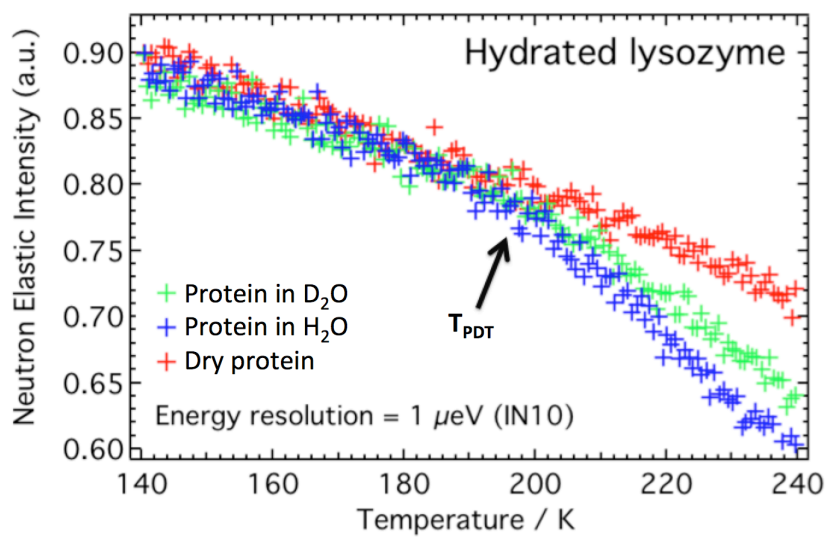

Fig. 2 – Total elastic neutron scattered intensity versus temperature for $D_2O$ hydrated lysozyme (green), $H_2O$ hydrated lysozyme (blue), and dry lysozyme (red) at 1μeV resolution collected on the IN10 neutron spectrometer at the Institut Laue-Langevin in Grenoble, France. This shows that (at this resolution) the dynamics of the protein and of its hydration water are almost indistinguishable, and consequently no evidence for de-coupling of the dynamics. However, comparison of hydrated and dry lysozyme clearly shows the coupling between the dynamics of the protein and of its hydration water, that starts to relax at $T_{PDT}$=200K.

On the other hand, there is a collection of experimental and computational results[20-26] that are indeed in apparent disagreement with the above-pictured PDT scenario[12-14,18,19,27]. Several of this out-of-scenario results are based on the fact that no changes in the dynamical behaviour of proteins seem to occur at all, but they do seem to agree on a sort of dynamical transition in the relaxation dynamics of the protein hydration-water. In summary, the PDT scenario is supported by the experimentally-measured coupling between the dynamics of proteins and their hydration water as shown in Fig. 2, whilst de-coupling suggested by other studies strongly suggests that some new scenario may actually apply.

We aim to solve this 30 year-old puzzle in protein hydration-water dynamics by using elastic neutron scattering at the highest achievable resolution for this type of study, that is 0.3 µeV in place of the more usual 1µeV used so far. We use the model protein, lysozyme. The rationale is that a relaxation process of the system may be too slow to be measurable at a given energy-resolution, or (in the present case) the relaxation-times of two or more processes are so similar that they cannot be separated at this given resolution. In either case, higher resolution is required. For more details about elastic neutron scattering spectroscopy and energy resolution refer to Ref. [28,29]. In Fig. 3 the elastic intensity as a function of temperature for dry and hydrated lysozyme is presented. The data have been collected on the IN16B spectrometer at the Institut Laue-Langevin in Grenoble, France in its highest resolution set-up, $\Delta\omega_{RES}$=0.3µeV. This highest energy resolution allows us to measure the de-coupling between the dynamics of proteins and their hydration water for the first time. All previous experiments on hydrated proteins have been done with worse energy resolutions (around 0.8µeV). Figure 3 clearly shows the de-coupling between the hydration-water motion and the protein-relaxation dynamics, and it follows that motion in the hydration water plays no significant role in driving any transition of the protein dynamics. Hydration water relaxes at $T_W$=179±1K, whereas protein relaxes at $T_P$=195±1K, i.e. 16K above (Fig. 3c). Thus, the experimental data in Fig. 3 refute the well-accepted PDT scenario presented earlier in this letter[12-14,18,19,27] that proposed the motions of hydration water trigger sudden changes in the protein dynamics, and hence function.

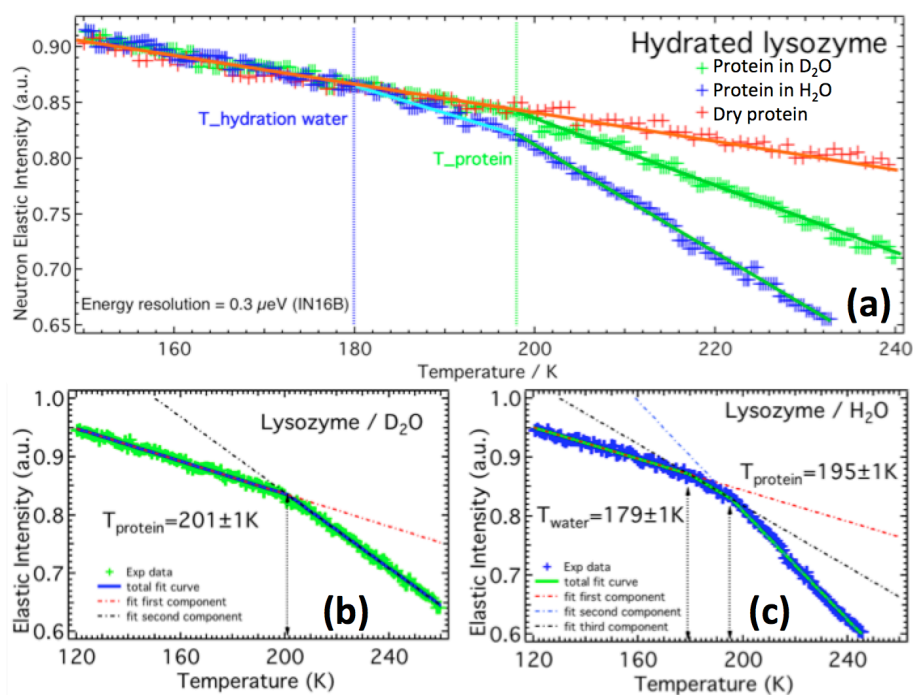

Fig. 3 – High-resolution elastic neutron data showing the de-coupling between the dynamics of protein and of its hydration water. Total elastic neutron-scattered intensity versus temperature for $D_2O$ hydrated lysozyme (green), $H_2O$ hydrated lysozyme (blue), and dry lysozyme (red). The comparison in (a) clearly shows the de-coupling between the dynamics of lysozyme and of their hydration water. The protein starts to relax at $T_P$=195K, de-coupled from the transition in the dynamics of hydration water, which only starts to relax 16K lower, i.e. $T_W$=179K. (b-c) The high counting-statistic of the experiment also allows the effect of hydrating with heavy water: in this case the protein dynamics is slightly slower that when hydrated in isotopically-normal water.

In conclusion, with the limited energy resolutions used in the past (up to 0.8μeV on the best high-resolution spectrometers), proteins and their hydration water appeared to undergo a dynamical transition at the same temperature. This inferred the erroneous conclusion of coupled water-protein relaxation dynamics, and, in turn, the conjecture that transitions in hydration-water dynamics are responsible for protein dynamics and function. Our high-resolution experimental data (Fig. 3) resolve this misconception, and reveal for the first time, that there is de-coupling between the relaxation of a model protein and the relaxation of its hydration water. This is an important milestone to this 30 year-old puzzle, and shows that relaxations in the hydration water dynamics do not induce any transition in protein dynamics.

**Acknowledgment:**
A.B. thanks Profs. Pietro Ballone and Gordon J. Kearley for fruitful discussions. A.B. acknowledges support from (i) the European Community under the Marie-Curie Fellowship Grants HYDRA (No. 301463) and PSI-FELLOW (No. 290605), and from (ii) Science Foundation Ireland (SFI) under the Start Investigator Research Grant 15-SIRG-3538, with additional support provided by the School of Physics, University College Dublin, Ireland, and the Laboratory for Neutron Scattering, Paul Scherrer Institute (PSI), Switzerland. A.B. acknowledges the Institut Laue-Langevin (Grenoble, France) for the access to IN10 and IN16B spectrometers, and Drs. Miguel Gonzales and Bernard Frick, in particular, for the help during the neutron beam-time. A.B. acknowledges Drs. Ekaterina Pomjakushina and Antonietta Gasperina (PSI) for the access to the differential scanning calorimetry laboratory, and the bio-lab, respectively.